*Chapter 11*

# Demand for shared mobility to replace private mobility using connected and automated vehicles

*Seyed Mehdi Meshkani [1] Shadi Djavadian[2] and Bilal Farooq[1]*

*In:*

## Shared Mobility and Automated Vehicles: Responding to socio-technical changes and pandemics

*Edited by: Ata Khan and Susan Shaheen*

We examine how introduction of Shared Connected and Automated vehicles (SCAVs) as a new mobility mode could affect travel demand, welfare, as well as traffic congestion in the network. To do so, we adapt an agent-based day-to-day adjustment process and develop a central dispatching system, which is implemented on an in-house traffic microsimulator. We consider a two-sided market in which demand and SCAV fleet size change endogenously. For dispatching SCAV fleet size, we take changing traffic conditions into account. There are two available transport modes: private Connected Automated Vehicles (CAVs) and SCAVs. The designed system is applied on downtown Toronto network using real data. The results show that demand of SCAVs goes up by 43 per cent over seven study days from 670 trips on the first day to 959 trips on the seventh day. Whereas, there is a 10 per cent reduction in private CAV demand from 2807 trips to 2518 trips during the same duration. Moreover, total travel time of the network goes down by seven per cent indicating that traffic congestion was reduced in the network.

[1] Laboratory of Innovations in Transportation (LiTrans), Department of Civil Engineering, Ryerson University

[2] Ford Mobility

## 11.1 Introduction

Due to increasing trends in urbanization, sustainable access to mobility has become a serious issue. The pace of development of services and infrastructures in cities has traditionally been slower than the pace of increase in transportation demand [1]. Existing solutions such as add parking spaces or expanding roadways have various consequences, including raising environmental concerns, threatening the livability of cities, and in many cases, being expensive. Therefore, it is necessary to consider new and potentially transformative transportation solutions [2]. On the other hand, we are at the cusp of the emergence of CAVs that can drive themselves [3]. As discussed in Chapter 10, CAVs have the ability to facilitate car-sharing and ridesharing behavior. CAVs have a strong potential to implement on-demand shared mobility systems with or without ridesharing [3, 4].

Despite the obvious advantages of ridesharing, in order to be widely adopted, it must be able to compete with one of the biggest advantages of private car usage: immediate access to door-to-door transportation [5]. Moreover, there are some barriers, including the requirement that itineraries and schedules have to be coordinated between participants and the lack of effective methods to encourage participation that have inhibited a wider adoption of ridesharing [6]. Technological advances in hardware and software, nevertheless, have facilitated the problem to be solved. GPS-enabled smartphones, social networks, data repositories, and the internet enable practical dynamic or realtime ridesharing. Dynamic ridesharing refers to a system which supports an automatic ride-matching process between participants on very short notice or even en-route [7, 8].

SCAVs either with or without ridesharing, enable people to schedule a ride using mobile phone applications. Moreover, the system can anticipate future short-term demand and relocate vehicles in advance to better match supply with travel demand. Environmental benefits, particularly in the form of reduced parking and vehicle ownership needs as well as the potential for additional vehicle-kilometer traveled (VKT) reduction are some benefits of using SCAVs [9, 10].

One question may rise here is that how would these emerging SCAVs with ridesharing capability as a new transportation mode affect demand of private mode as well as users' welfare. Another question is that how they would influence traffic congestion on the network. As discussed in chapter 10, a central operated CAV fleet in comparison with crowdsourced HDV (e.g. current version of Uber and Lyft) fleet yields better results and stability. In this study, as an extension of chapter 10, we answer aforementioned questions and evaluate the effects of SCAVs as a new mode on travel demand, users' welfare and traffic congestion. To do so, we employ an agent-based simulation framework for SCAVs. We develop a central dispatcher as well as SCAV demand and SCAV fleet size estimators. Demand and SCAV fleet size change every day based on day-to-day adjustment process developed by Djavadian and Chow [11, 12]. The designed SCAV system is implemented and tested on an in-house agent-based traffic microsimulator developed by Djavadian and Farooq [13].

The remainder of this paper is organized as follows. A general review of literature related to shared mobility will be presented in the background section. In the methodology section, we explain the designed framework for SCAVs. The case study section illustrates the performance of the designed system on a real instance network, and finally, in the conclusion, results and future works are presented.

## 11.2 Background

Shared mobility (e.g. ride hailing and ridesharing) has gained significant attention in the literature over the recent years. However, to the best of authors' knowledge, most of the previous literature assessed the operations of human driven vehicles (HDV) either from the operator or users' perspective. Recently, several research efforts have been investigated to evaluate the performance of using SCAV fleet on urban transportation network. SCAV fleet size, wait time, financial and environmental impacts are some indicators that have been studied. One of the most important potentials of SCAVs is the capability to rideshare. Ridesharing is a transportation mode which benefits both users and society by saving travel cost, reducing travel time, mitigating traffic congestions, conserving fuel, and reducing air pollution.

Burns et al. [14] examined the performance of a shared automated fleet in three distinct city environments: a mid-sized city (Ann Arbor, Michigan), a low-density suburban development (Babcock Ranch, Florida), and a large densely-populated urban area (Manhattan, New York). They compared the SCAVs with privately owned ones in terms of operational cost and fleet size. The study found that in mid-sized urban and suburban settings, each shared vehicle could replace 6.7 privately owned vehicles. Meanwhile, in the dense urban setting, the current taxi fleet could be downsized by 30 per cent with the introduction of automated driving technology with average wait times at less than one min.

Fagnant and Kockelman [9] presented an agent-based system which examines the implications of SCAVs with capability of dynamic ride sharing across the Austin, Texas network. Results reveal that dynamic ridesharing leads to a reduction in total service times (wait times plus in-vehicle travel times) and travel costs for SCAV users. In addition, overall vehicle-miles traveled can be reduced as trip-making intensity (SCAV membership) rises and/or users become more flexible in their trip timing and routing. Zhang et al. [3] using an agent-based simulation model in a grid based hypothetical city, estimated how SCAVs affect urban parking demand under different operations scenarios. They reported that one SCAV will be able to replace around 14 privately owned vehicles and approximately 90 per cent of parking demand for the participating clients can be reduced. In another research effort, Chen et al. [15] examined the implications of shared, automated, electric vehicles fleet in a discrete time agent-based model. The simulation was conducted under various vehicle range and charging infrastructure scenarios in a 100-mile by 100-mile gridded city of Austin, Texas. Results reveal that fleet size is sensitive to battery recharge time and vehicle range. Simulation predicts a decrease in fleet empty vehicle miles (3 to 4 per cent) and average wait times (2 to 4 minutes per trip), with each vehicle replacing 5 to 9 privately owned vehicles.

Levin et al. [16] proposed an event-based framework in which Cell Transmission Model (CTM) was used as a realistic flow model to obtain more accurate predictions about SCAV operations. A heuristic was presented to study SCAVs with dynamic ridesharing capability. The authors compared SCAVs scenarios (including dynamic ridesharing) with personal vehicle scenarios. Results reveal that a smaller SCAV fleet can service all travel demand in the AM peak.

However, some SCAV scenarios also increased congestion due to empty repositioning trips to reach travelers' origins. In such a situation, it is important to model congestion when studying SCAVs to attain realistic estimates of quality of service. Furthermore, SCAVs may be less effective than previously predicted for peak-hour scenarios. Nevertheless, SCAVs with dynamic ridesharing provided service comparable to personal vehicles.

Becker et al. [17] developed the first joint simulation of carsharing, bikesharing and ridehailing for a city-scale transport system with a fixed fleet size using MATSim. Their results showed that introduction of shared modes may increase transport system efficiency by up to seven per cent and this efficiency may reach to 11 per cent if share modes were used as a substitute for public transport in lower-density areas.

Most of above-mentioned studies have considered one-sided market, i.e. either demand or fleet size is fixed as an exogenous input. There are only few previous studies e.g. Djavadian and Chow [12] or Chapter 10 that assessed a two-sided market using an agent based day-to-day adjustment simulation in which both demand and fleet size are determined endogenously. However, in their proposed framework for dynamic on demand services, they did not consider traffic congestion in the network. This study in terms of demand and supply considers a two sided-market. In addition, for simulating SCAVs, traffic congestion in the network is taken into account. Note that in chapter 10, for ridesharing services just HDVs and CAVs has been considered. Whereas, in this study we consider private CAVs and SCAVs as the available modes. Another difference is that chapter 10 addressed the first mile/last mile problem which is considered many-to-one problem while this study addresses many-to-many.

## 11.3 Methodology

In this study, we employ an agent-based simulation framework for modeling SCAVs. As can been seen in **Figure 11.1,** the framework is composed of three major parts that have interaction with each other: 1) dynamic demand, including demand for private CAVs and demand for SCAVs 2) centralized SCAV system which consists of SCAV demand, SCAV fleet size and dispatcher and 3) an in-house traffic microsimulator, which model traffic conditions in the network.

**Figure 11.1 ABOUT HERE".**

*Figure 11.1 Agent based Simulation Framework for SCAVs*

### *11.3.1 Dynamic Demand*

Users in the network choose their mode of transportation based on a learning process gained from their experiences from the previous days. There are two available modes in the network: private CAVs and SCAVs. To model users' mode choice a binary logit model is developed which shows the choice of users on each day. The number of users choosing each mode creates the demand of private CAVs and the demand of SCAVs. This demand changes day-to-day, as on each day some users might switch from CAV to SCAV and vice versa. Eventually, after several days of learning, they would reach to a steady state. For the day-to-day adjustment process, we use similar method developed by Djavadian and Chow [11, 12].

### *11.3.2 Centralized SCAV Dispatcher System*

The centralized SCAV system proposed in this study consists of two network layers and three types of agents. The two network layers are: communication network where the information exchange between vehicles, links and intersections takes place, and the physical road network where vehicles travel on. Physical road network is represented by a network $G(I, L)$ which consists of $I$ intersections (nodes/vertices) and $L$ links (edges). Three agents in this system are SCAV agents ($f \in F$), passenger agents ($p \in P$), and infrastructure agents. there is a central dispatcher ($D$) in the system which is responsible for receiving requests from travelers, matching them to SCAVs and ensuring all travelers are served. However, no central depot exists and depots are distributed throughout the network. $\Delta$ is dispatch update cycle which is set to 1 min in this study, and $j$ is time whose unit is second.

This system is based on several assumptions: a) only intersections where produce demand has depot b) SCAVs' capacity is four passengers c) There is no threshold for waiting time and passengers always wait on the waitlist.

Once a passenger needs a SCAV, they make a request to the dispatcher. Dispatcher checks to see if there is any SCAV available at intersection depot where the passenger makes request. If available, assigns it to the passenger, if SCAV is not available, the passenger request would place on the waitlist. Dispatcher is aware of the status of all SCAVs, including location as well as capacity. When any SCAV arrives at this intersection, dispatcher checks to see if there is enough capacity and then assigned it to the passenger waiting at this intersection. If there are multiple requests at intersection, passengers are serviced in a first-come-first-serve (FCFS) order.

When a SCAV is assigned to a passenger, SCAV picks him up from the intersection, on en-route to passengers' destination, while arriving at each intersection first check to see if any passenger needs to be dropped off then check to see if dispatcher assigns SCAV any new passenger. The general policy for dropping off is first- in first- out (FIFO). Each SCAV after dropping off the last passenger goes back to its own depot, if en-route to the depot there is not any new assignment. The pseudo-code for the proposed centralized system is as **Figure 11.2**.

**Figure 11.2 ABOUT HERE".**

*Figure 11.2 Pseudo-code for Centralized SCAV Dispatcher System*

Similar to the SCAV demand discussed in the previous part, SCAV dispatcher has a learning process and is updated day-to-day using **Equation 10.3** from Chapter 10.

### *11.3.3 Traffic Microsimulator*

The proposed centralized SCAV system is implemented on an in-house traffic microsimulator which already developed by Djavadian and Farooq [13]. In this simulator, a traffic management system using network of intelligent intersections was proposed that is capable of dynamically routing intelligent vehicles (connected and automated) from origin to destination in dense urban areas. In this system, realtime traffic information is collected by links using sensors and being frequently exchanged among intelligent intersection using Infrastructure to Infrastructure (I2I) communication.

## 11.4 Case study

To test the framework on real-life instances, it is implemented for downtown Toronto road network. Because of the high current demand levels and congestion in downtown Toronto network, it is a good test case for the proposed algorithm. **Figure 11.3** presents the test network, which consists of 76 nodes, 223 links and 26 centroids (matched to nearest nodes). As mentioned in the methodology section, depots are distributed throughout the network. Study period for our simulation is 7:45am-8:00am which is part of morning peak period and just vehicles whose both origin and destinations are in downtown Toronto are considered. The demand used in this study is time dependent exogenous demand Origin-Destination (OD) matrices which is based on 5mins intervals from the 2011 Transportation Tomorrow Survey (TTS) [18]. We used growth factor to convert this data to today's demand. During the study period, total demand is 3477 on the test network which is distributed randomly within 5mins using a Poisson distribution. The simulation is run for seven consecutive days for the demand and supply to come to a steady state.

**Figure 11.3 ABOUT HERE".**

*Figure 11.3 Downtown Toronto Street Network*

The objectives of this case study are to evaluate how introducing a shared transport mode using SCAVs affects users' mode choice as well as traffic congestion. To model users' mode choice, we developed a binary logit model which represents in **Table 11.1** and **Table 11.2**. We synthesize a population such that all observable traits are captured from survey data and all unobservable variables $\varepsilon$ are randomly drawn for each user $n$ to fit the observed choices from sample data [11].

*Table 11.1 Utility Function*

| # | Mode | Utility |
|---|---|---|
| 1 | CAV (Car) | $U_{(CAV)} = ASC_{CAV} + \beta_1(T_{CAV} - T_{SCAV}) + \beta_2(ratio^*) + \varepsilon$ |
| 2 | SCAV | $U_{(SCAV)} = 0$ |

$$* \ ratio = \frac{\# \ of \ cars \ in \ houshold}{\# \ of \ licence \ in \ the \ houselohd}$$

Where $T_{CAV}$ is total travel time of private CAV, and $T_{SCAV}$ is total travel time of SCAV, including wait time and in-vehicle time.

*Table 11.2 Utility Parameters*

| Name | Value | std err | t-test |
|---|---|---|---|
| $ASC_{car}$ | $-1.91$ | 1.22 | $-1.56$ |
| $\beta_1$ | $-0.153$ | 0.0798 | $-1.92$ |
| $\beta_2$ | 2.24 | 0.776 | 2.89 |
| Initial log-likelihood | -42.975 | | |
| Final log-likelihood | -21.038 | | |
| Likelihood ratio test | 43.874 | | |
| Adjusted rho-square | 0.441 | | |

Two scenarios are considered in this study. The base case scenario represents the demand for private CAV and current public transit as two available modes. In the scenario # 1, we replace the public transit with on-demand SCAV and run it for seven consecutive days and then compare the demand of each day with base case scenario. In the base case, from the overall 3,477 demand, 2,807 choose CAV and 670 choose public transit. The current demand of public transit is used for the first day. In the scenario # 1, we set the initial number of SCAV fleet 200 and maximum number of 300. A learning process is used for updating demand and fleet size each day based on Djavadian and Chow [11, 12].

## 11.5 Results

In this study, different indicators, including CAV and SCAV demand, SCAV fleet size, welfare level, and network total travel time for seven consecutive days were measured. **Figure 11.4** illustrates the demand for CAV and SCAV over these seven days and **Figure 11.5** shows the SCAV fleet size. As can be seen in **Figure 11.4**, the demand of public transit in the base case is 670 trips which by replacing SCAV, more users switch to SCAV and this value increases to 959 trips on the seventh day showing 43 per cent growth. In contrast, CAV demand decreases from 2807 trips to 2518 which is 10 per cent reduction and shows that users switch from private CAVs to SCAVs. This switch to SCAV leads to 19 per cent increase in SCAV fleet size from 200 vehicles on the first day to 238 vehicles on the seventh day which can be seen in **Figure 11.5.** Based on the **Figure 11.4**, for the second day, SCAV demand increases whereas CAV reduces; this is because on the first day, SCAV users get a good experience of using SCAVs in terms of total travel time, so, on the second day, more users are willing to use SCAVs. On the other hand, since SCAV fleet size is determined based on SCAV demand from the previous day, on the second day, there is a reduction in the fleet size, which shown in **Figure 11.5,** whereas SCAV demand grows. Therefore, SCAV users on the second day experience some inconvenience in terms of total travel time. That's why for the third day in **Figure 11.4** there is a reduction in SCAV and a rise in CAV demand. This interaction between demand and SCAV fleet size is done for all other days to reach an equilibrium condition in which users don't change their transport mode and demand of each mode for two consecutive days are the same.

**"Figure 11.4 ABOUT HERE".**

*Figure 11.4 CAV and SCAV demand*

**"Figure 11.5 ABOUT HERE".**

*Figure 11.5 SCAV Fleet Size*

**Figure 11.6** presents the total utility for users who used CAVs (cars). As can be seen, by introducing the SCAVs, total utility of the users who used CAVs decreases which means that in the presence of Shared CAVs, private CAVs is less attractive for users.

**"Figure 11.6 ABOUT HERE".**

*Figure 11.6 Total Users' Utility for CAV*

Another important measure that reflects on the congestion level in the network is the total travel time. As shown in **Figure 11.7**, introduction of a SCAV mode with ridesharing result in decrease in total travel time from 35675 (veh.min) on the first day to 33220 (veh.min) on the seventh day which shows a seven per cent reduction in total travel time and consequently indicates that traffic congestion decreases in the network.

**"Figure 11.7 ABOUT HERE".**

*Figure 11.7 Network Total Travel Time*

## 11.6 Conclusion

In this study, we analyzed SCAVs as a new mode to replace private vehicles in the future. To do so, we employed an agent-based simulation framework, which is composed of dynamic demand, SCAV system, and an in-house traffic microsimulator. For SCAV system, we developed a centralized dispatcher for managing SCAVs as well as passengers in the network. Dynamic demand and SCAV fleet size are updated every day using a day-to-day adjustment process in a two-sided market [11, 12]. The in-house traffic microsimulator [13] routes CAVs in the network based on dynamic traffic conditions. In our framework, demand and SCAV fleet size are determined each day endogenously. For simulating shared CAVs, we considered traffic congestion in the network. Two research questions are answered in this study: How introduction of a SCAVs as a new shared mode affects demand of private cars and how it influences traffic congestion in the network. Each day, people choose their mode of transport based on their experience from previous day. To capture individual's mode choice decisions, a binary logit model was developed. Downtown Toronto network is chosen to implement the proposed system with real data for seven consecutive days. Results show that the demand of SCAVs goes up from 670 trips on the first day to 959 trips on the seventh day showing 43 per cent growth over these seven days whereas CAVs demand drops from 2807 trips to 2518 indicating 10 per cent decrease which reflects the efficiency of shared mobility mode. SCAV fleet size goes up by 19 per cent from 200 vehicles to 238 vehicles over these seven days, in order to meet the extra demand of those who change their mode to SCAVs. Total network travel time is also measured and shows a 7 per cent reduction over the seven days which indicates that using SCAVs as a shared mode has a positive effect on traffic congestion of the network.

A number of directions can be taken in future research. This study focuses on centralized dispatching system, while in the future distributed dispatching systems can be addressed. In the proposed dispatching system in this study, passengers are serviced based on first in first out (FIFO) while in the future, policy can be made based on advanced optimization models with differentiated service. Another direction is that relocation can be taken into account, since it reduces the vehicle kilometer traveled in the system.

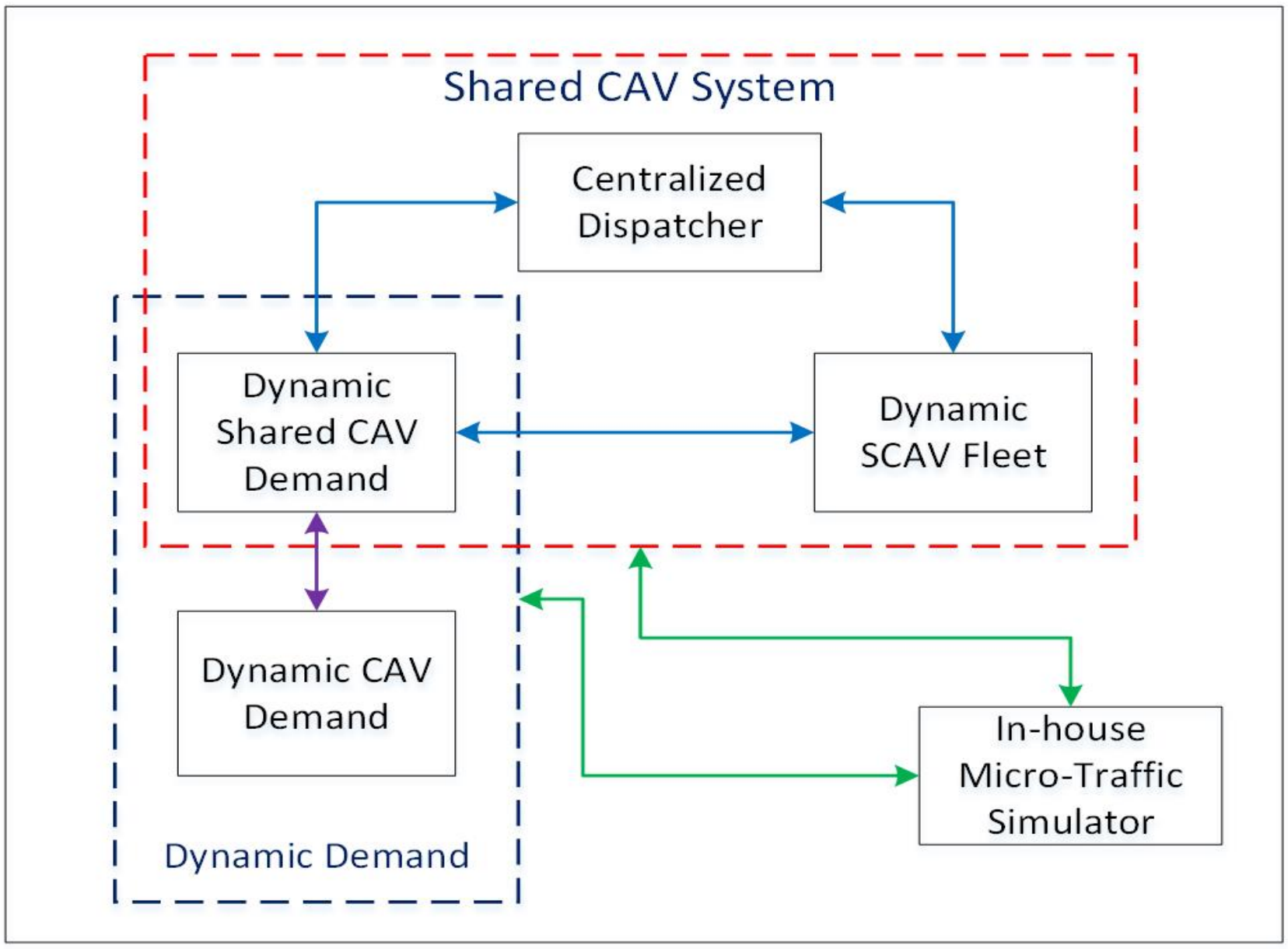

*Figure 11.1 Agent-based Simulation Framework for SCAVs*

```
j=1;
while # of passengers < total demand do
    if (j mod Δ_j) = 0 then
        ... Dispatcher
        Passenger (p) sends a request for SCAV (f) to the dispatcher;
        while p <= total number of requests do
            Dispatcher creates a request-table based on passengers' itinerary;
            Dispatcher checks to see if there is available SCAV at intersection depot;
            if SCAV (f) is available and has enough capacity then
                Dispatcher assigns to passenger (p);
                Dispatcher updates request table, including passenger status to Assigned and SCAV
                  status to busy;
                SCAV (f) creates an itinerary for on board passengers;
                if SCAV (f) does not have any capacity then
                    leaves intersection I_i;
                end
            else
                Passenger (p) waits on request-table for a passing SCAV from intersection (I_i) or a
                  SCAV coming back to intersection I_i;
            end
            ... Vehicle trajectory update
            while f<=F do
                When SCAV arrives at I_i Checks to see if any passenger needs to be dropped off from
                  customer on board itinerary or picked up from request table;
                If picking up or dropping off, dispatcher and f update their request and itinerary
                  tables;
                If all passengers of f have been dropped off, it goes back to the depot; f=f+1 ;
            end
        end
    end
    j=j+1;
end
```

*Figure 11.2 Pseudo-code for Centralized SCAV Dispatcher System*

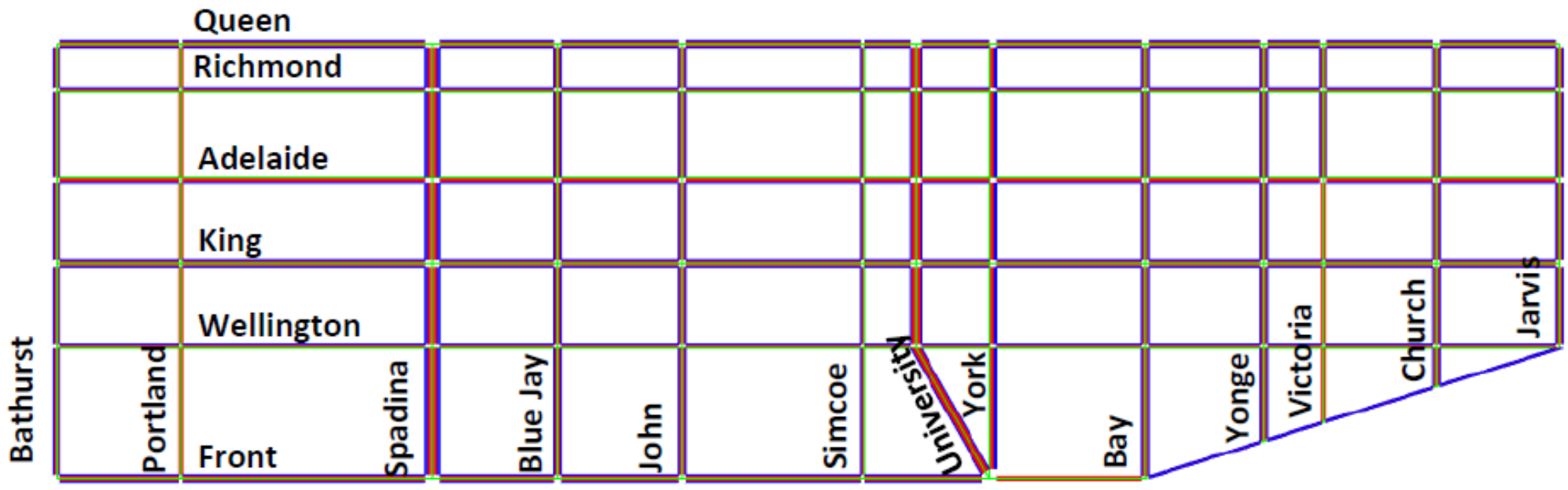

*Figure 11.3 Downtown Toronto Street Network* [13]

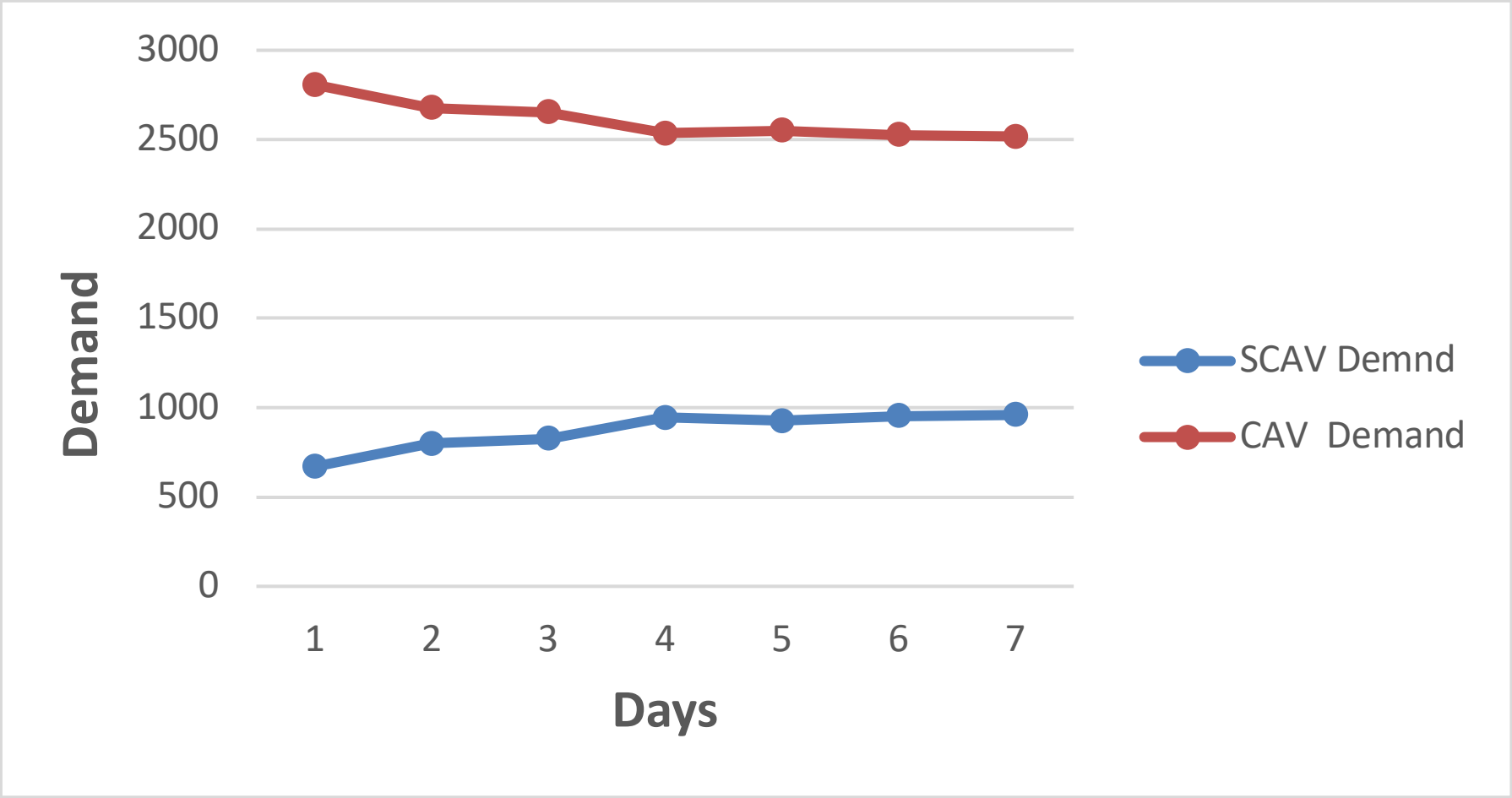

*Figure 11.4 CAV and SCAV demand*

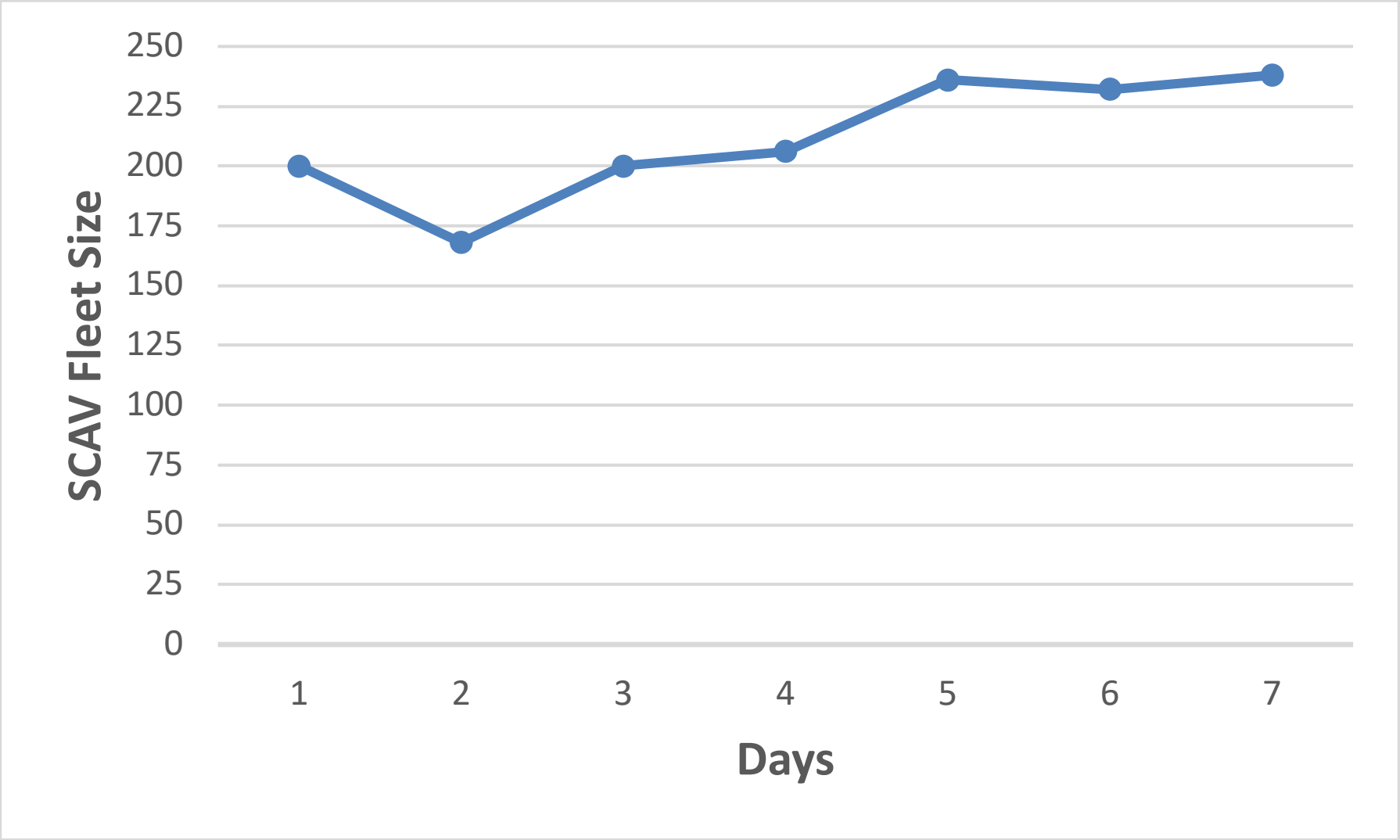

*Figure 11.5 SCAV Fleet Size*

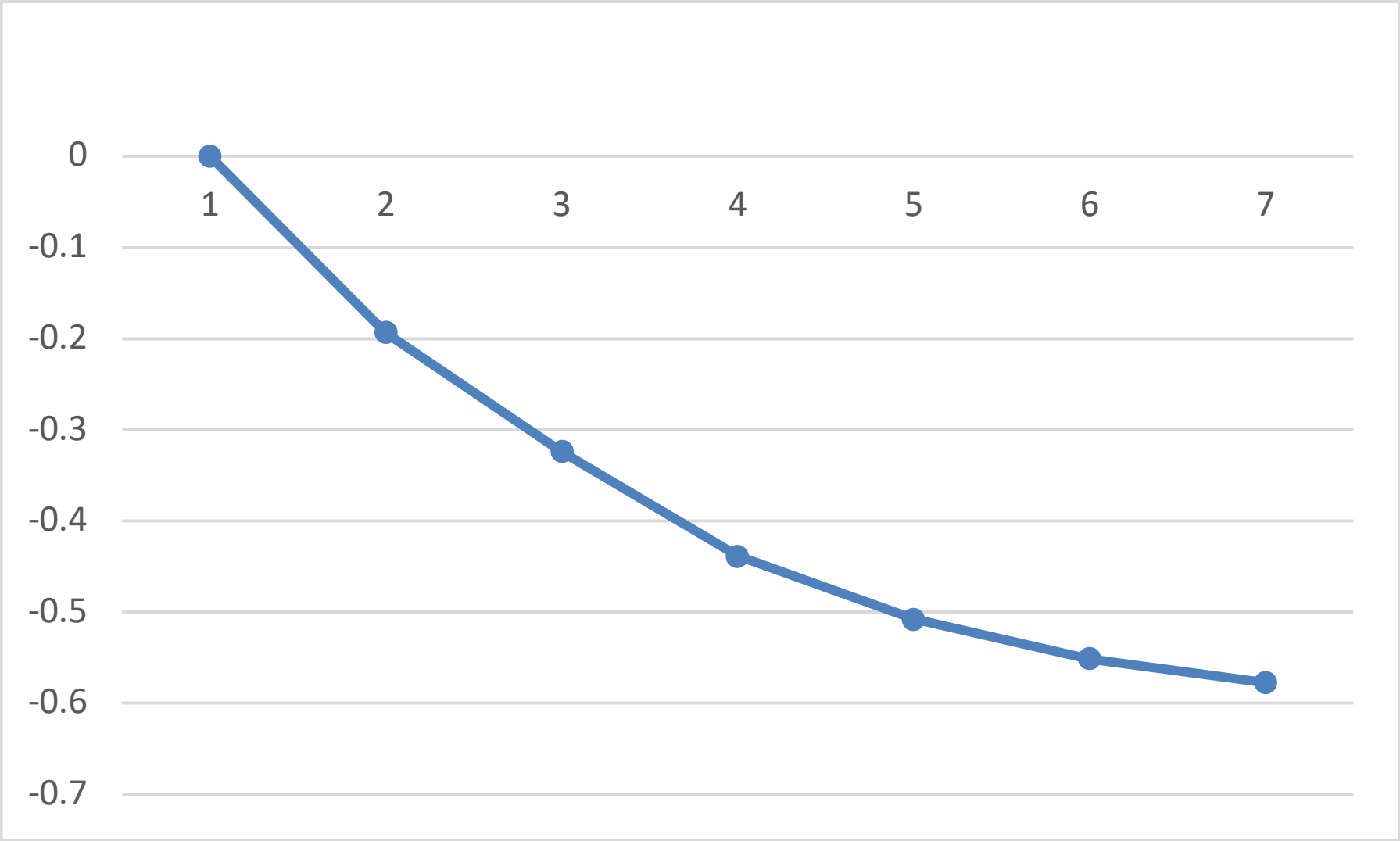

*Figure 11.6 Normalized Total Users' Utility for CAVs*

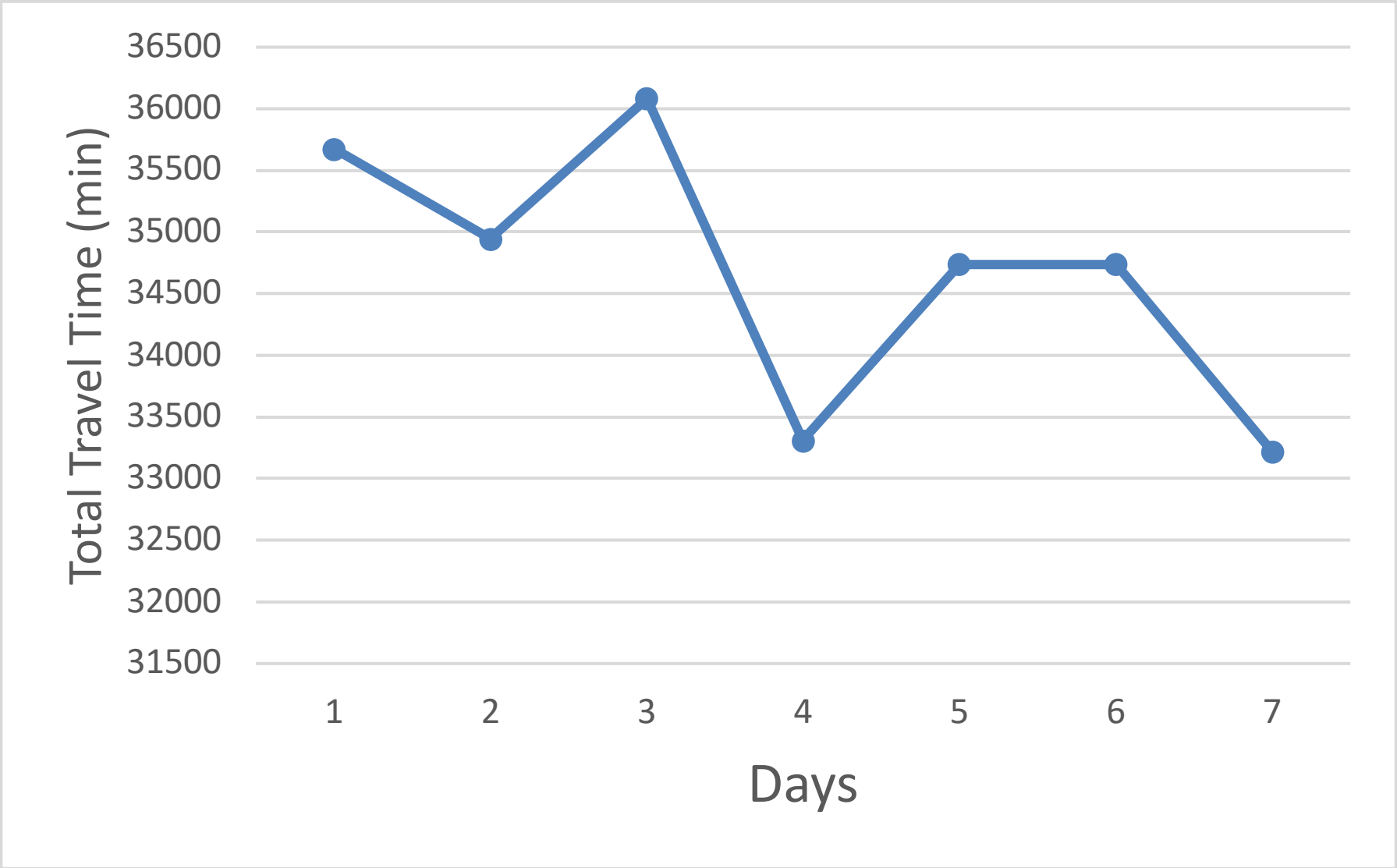

*Figure 11.7 Network Total Travel Time*